\documentclass[journal ]{new-aiaa}
\usepackage[utf8]{inputenc}
\usepackage{textcomp}
\usepackage{graphicx}
\usepackage[version=4]{mhchem}
\usepackage{siunitx}
\usepackage{longtable,tabularx}
\usepackage[nolist,nohyperlinks]{acronym}
\usepackage{booktabs} %
\usepackage{float} %
\usepackage{amsmath, bm,bbm,latexsym,subcaption,epstopdf}
\usepackage{fix-cm} %
\usepackage{caption} %
\usepackage{subcaption} %
\usepackage{tikz} %
\usetikzlibrary{positioning, calc, arrows, shapes, decorations}
\usepackage{multirow}
\usepackage{placeins}    %

\graphicspath{{../figures/}}
\DeclareGraphicsExtensions{.pdf,.eps}

\begin{acronym}[XXXXXXX] %
\acro{ATR}{automatic target recognition}\label{ATR}
\acro{BTT}{bank-to-turn}\label{BTT}
\acro{COTS}{commercial off-the-shelf}\label{COTS}
\acro{CRLB}{Cramér-Rao lower bound}\label{CRLB}
\acro{EKF}{extended Kalman filter}\label{EKF}
\acro{EO}{electro-optical}\label{EO}
\acro{EOM}{equations of motion}\label{EOM}
\acro{FIM}{Fisher information matrix}\label{FIM}
\acro{FOV}{field-of-view}\label{FOV}
\acro{FFOV}{fixed field-of-view}\label{FFOV}
\acro{FKF}{federated Kalman filter}\label{FKF}
\acro{GPS}{global positioning system}\label{GPS}
\acro{INS}{inertial navigation system}\label{INS}
\acro{LOS}{line-of-sight}\label{LOS}
\acro{NED}{north-east-down}\label{NED}
\acro{NFZ}{no-fly zone}\label{NFZ}
\acro{NIIRS}{National Imagery Interpretability Rating Scale}\label{NIIRS}
\acro{NLP}{nonlinear programming}\label{NLP}
\acro{OTH}{over-the-horizon}\label{OTH}
\acro{PCRLB}{posterior Cramér–Rao lower bound}\label{PCRLB}
\acro{PDOP}{position dilution of precision}\label{PDOP}
\acro{pixel}{picture element}\label{pixel}
\acro{PS}{Pseudospectral}\label{PS}
\acro{resel}{resolvable element}\label{resel}
\acro{RMSE}{root mean square error}\label{RMSE}
\acro{SQP}{sequential quadratic programming}\label{SQP}
\acro{SWaP-C}{size, weight, power, and cost}\label{SWaP-C}
\acro{UAV}{unmanned aerial vehicle}\label{UAV}
\acro{USV}{unmanned surface vessel}\label{USV}
\end{acronym}

    \tikzset{
        block/.style = {draw, rectangle,
            minimum height=1cm,
            minimum width=2cm},
        input/.style = {coordinate,node distance=1cm},
        output/.style = {coordinate,node distance=4cm},
        arrow/.style={draw, -latex,node distance=2cm},
        pinstyle/.style = {pin edge={latex-, black,node distance=2cm}},
        sum/.style = {draw, circle, node distance=1cm},
    }

\title{Multi-UAV Trajectory Optimization for Bearing-Only Localization in GPS-Denied Environments}

 \author{Alfonso Sciacchitano\footnote{Former Doctoral Student, Dept. of Mechanical and Aerospace Engineering; currently Project Officer, Supervisor of Shipbuilding, Conversion \& Repair, Newport News, VA. Email: \texttt{alfonso.sciacchitano@gmail.com}, Member AIAA (Corresponding Author).},
 Liraz Mudrik\footnote{Postdoctoral Fellow, Dept. of Mechanical and Aerospace Engineering, Email: \texttt{liraz109@gmail.com}, Member AIAA.},
  	Sean Kragelund\footnote{Research Associate Professor, Dept. of Mechanical and Aerospace Engineering, Email: \texttt{spkragel@nps.edu}, Member AIAA.},
 	and  
 	Isaac Kaminer\footnote{Professor, Dept. of Mechanical and Aerospace Engineering. Email: \texttt{kaminer@nps.edu}, Senior Member AIAA.
    \newline\newline Partially presented as Paper AIAA 2025-2604 at
    the AIAA SciTech 2025 Forum, Orlando, FL, 6--10 January
    2025. This manuscript was published open access in the
    Journal of Guidance, Control, and Dynamics,
    \url{https://doi.org/10.2514/1.G009750}; please read and cite
    the published version. This material is
    declared a work of the U.S. Government and is not subject to
    copyright protection in the United States.}}	
\affil{Naval Postgraduate School, Monterey, CA, 93943}

\begin{document}

\maketitle
\setcounter{footnote}{4} 

\begin{abstract}
Accurate localization of maritime targets by unmanned aerial vehicles (UAVs) remains challenging in GPS-denied environments. UAVs equipped with gimballed electro-optical sensors are typically used to localize targets; however, reliance on these sensors increases mechanical complexity, cost, and susceptibility to single-point failures, limiting scalability and robustness in multi-UAV operations. This work presents a new trajectory optimization framework that enables cooperative target localization using UAVs with fixed, non-gimballed cameras operating in coordination with a surface vessel. This estimation-aware optimization generates dynamically feasible trajectories that explicitly account for mission constraints, platform dynamics, and out-of-frame events. Estimation-aware trajectories outperform heuristic paths by reducing localization error by more than a factor of two, motivating their use in cooperative operations. Results further demonstrate that coordinated UAVs with fixed, non-gimballed cameras achieve localization accuracy that meets or exceeds that of single gimballed systems, while substantially lowering system complexity and cost, enabling scalability, and enhancing mission resilience.
\end{abstract}
 
\section{Introduction}\label{Sec:Introduction}
Cooperative operations between \acfp{UAV} and \acfp{USV} enhance situational awareness, extend sensing range, and enable autonomous engagement in denied or contested environments~\cite{Telli_2023}. Among these operations, target localization is mission-critical, demanding solutions that are accurate, robust, responsive, and cost-effective~\cite{Krystosik_2021}. Traditional methods, such as \acf{GPS}-based tracking and active ranging (e.g., radar, LiDAR), perform well in permissive settings but degrade significantly under \ac{GPS}-denied or electromagnetically contested conditions. Specifically, GPS can be jammed or spoofed, and active ranging sensors are both vulnerable to electronic countermeasures and risk revealing the platform’s location through their own emissions. These limitations render them ill-suited for low-signature, long-endurance missions.

Passive alternatives, notably bearing-only measurements obtained from camera systems, address these challenges by using low-power \acf{EO} sensors to estimate target location through bearing measurements. Unlike active ranging systems, passive sensors do not emit signals, which reduces power consumption, lowers system cost, and enhances survivability by minimizing the risk of detection in contested environments. Additionally, many cooperative scenarios, such as UAV-assisted USV targeting of surface vessels, rely on relative geometry rather than absolute positioning, making passive sensing especially effective in denied environments~\cite[pp.~845--847]{Chung_2018}. 

Small fixed-wing \acp{UAV} equipped with passive \ac{EO} sensors are well-suited for these applications. Their mechanically simple design and minimal power requirements contribute to a favorable \acf{SWaP-C} profile, making them viable for endurance-limited, low-signature missions where stealth and agility are required, while also reducing cost, a critical factor given the high attrition rates expected in contested environments. The case for small fixed-wing UAVs is further underscored by recent operational losses of medium-altitude long-endurance platforms, such as the MQ-9 Reaper, with each exceeding \$30 million (USD) in unit cost~\cite{Peck}. Although these larger systems provide continuous target surveillance, their vulnerability in contested airspace significantly impacts overall mission success, as they are single points of failure. In contrast, coordinated low-cost UAVs offer resilient and economically sustainable alternatives for this mission type, as recent drone utilization has shown sortie costs of expendable low-altitude short-range \acp{UAV} to be several thousand dollars or less~\cite{Hambling_2025}.

In this work, a group of \acp{UAV} estimate the positions of both an adversarial surface target and a collaborating \ac{USV} using bearing-only measurements in a common local inertial frame established for the mission area. The \acp{UAV} are assumed to know their own states in this local frame through onboard navigation, initialized  relative to the collaborating \ac{USV}, rather than through global positioning such as \ac{GPS}. The collaborating \ac{USV} likewise does not rely on \ac{GPS} to determine its own absolute position during the engagement. Instead, the \ac{USV} relies on state estimates generated by the \acp{UAV} to recover the relative geometry between itself and the adversarial target within that local frame. %

\begin{figure}[!htb]
\centering
\includegraphics[width=0.6\linewidth]{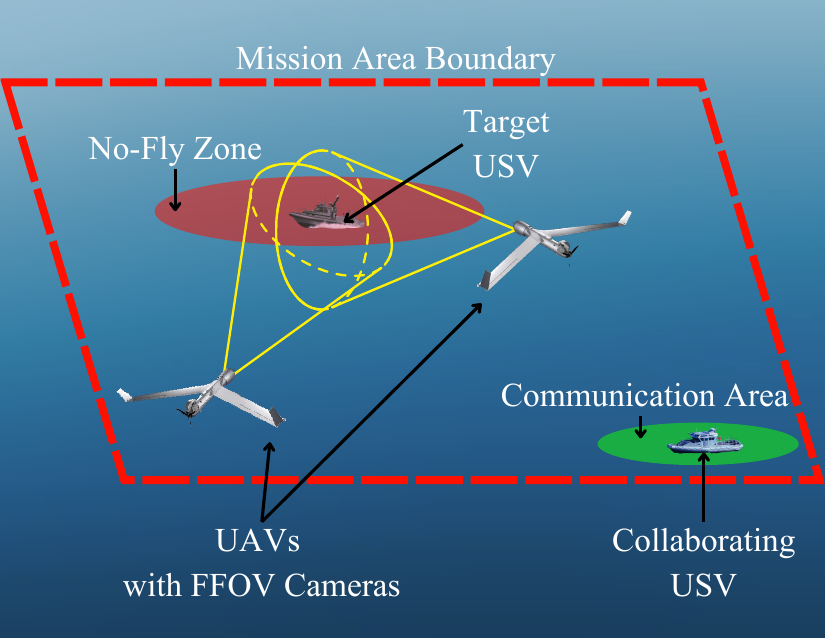}
\caption{Operational concept for cooperative localization using fixed-wing UAVs with fixed-field-of-view cameras. Local estimates are fused aboard the collaborating USV to determine relative target geometry.}
\label{fig:CONOPS}
\end{figure}

As illustrated in Figure~\ref{fig:CONOPS}, each \ac{UAV} acquires bearing-only measurements of both the adversarial target and the collaborating \ac{USV}. These measurements are processed onboard the \ac{UAV} to produce local state estimates, which are then relayed to the collaborating \ac{USV} at the conclusion of flight. Upon receiving these estimates, the \ac{USV} fuses them within a master filter that determines the a posteriori relative geometry between itself and the adversarial target. The objective of this work is therefore not to solve the full terminal guidance or interception problem for the collaborating \ac{USV}, but rather to provide the relative localization information needed to support subsequent targeting, navigation, or intercept actions by the collaborating platform.%

\subsection{Operational and Dynamic Constraints}
Prior studies on optimal observer motion for localization~\cite{hammel_optimal_1989,oshman_optimization_1999, ponda_trajectory_2009} prioritized estimation performance but omitted operational limitations such as sensor \ac{FOV}, mission‐area boundaries, and stealth requirements. Among prior studies, only a limited number have incorporated practical mission constraints. For example,~\cite{hammel_optimal_1989} considered a minimum‐time constraint, and~\cite{vander_hook_algorithms_2015} introduced communication‐distance limits. Consequently, these existing localization formulations lack the necessary fidelity to remain viable under realistic mission conditions.

Addressing this gap requires modeling \ac{EO} sensor \ac{FOV} restrictions which necessitate trajectory planning to optimize measurements while leveraging Bayesian estimation methods while the targets are out-of-frame, as described in~\cite[Chap.~12]{ristic_beyond_2003}. Additionally, trajectories must account for \ac{BTT} UAV dynamics, a characteristic of fixed-wing UAVs that rely on coordinated turns in which roll commands produce heading changes~\cite{Yokoyama_Ochi_2009}. This coupling between roll and yaw limits the UAV’s maneuverability and constrains the achievable flight paths, especially for \ac{FFOV} sensors that cannot pivot independently of the airframe. Consequently, the optimization must explicitly consider BTT dynamics to ensure that planned trajectories remain dynamically feasible and that \ac{EO} sensor \ac{LOS} constraints are satisfied throughout the mission. 

Additional constraints are imposed to maintain stealth and survivability, including \acp{NFZ} around adversarial targets, bounded UAV–USV communication distances to minimize detection risk and energy usage, and adherence to operational airspace limits such as altitude ceilings and geographic boundaries~\cite{Chai_2020}.
By explicitly embedding these constraints, the present study extends existing localization frameworks to generate operationally realistic paths that enhance UAV survivability and maximize overall mission effectiveness.

\subsection{Performance Metrics for Trajectory Optimization}
A widely adopted metric for improving estimation performance is the \ac{FIM}, whose inverse yields the \acf{CRLB}, a fundamental lower bound on estimation error~\cite{VanTrees_2013, ristic_beyond_2003}. In two-dimensional formulations, many researchers have chosen to maximize the determinant of the \ac{FIM} to generate trajectories that yield information-rich measurements~\cite{hammel_optimal_1989,passerieux_optimal_1998,oshman_optimization_1999,ousingsawat_optimal_2007,lee_cooperative_2013}. 

In three-dimensional target localization,~\cite{ponda_trajectory_2009} demonstrated that minimizing the trace of the \ac{CRLB} is more robust and computationally stable while leading to faster convergence than maximizing the determinant of the \ac{FIM}. Related marine robotics work has likewise adopted the trace of the inverse \ac{FIM} as the localization criterion for constrained three-dimensional bearings-only sensor placement problems and has shown that determinant-based criteria can be misleading in such settings~\cite{Moreno-Salinas_Pascoal_Aranda_2012,Moreno-Salinas_Pascoal_Aranda_2013}.

These attributes are particularly valuable in trajectory optimization for bearing-only localization, where measurements are strongly influenced by sensor geometry and platform constraints. Conceptually, the trace of the CRLB is analogous to \ac{PDOP} in GPS, where lower \ac{PDOP} values indicate more favorable satellite–receiver geometry and thus higher positioning accuracy. 

Similarly, minimizing the trace of the \ac{CRLB} favors trajectories that reduce aggregate estimation uncertainty by penalizing the sum of the state variances directly, rather than emphasizing only the overall information volume as in determinant-based \ac{FIM} criteria. This distinction is especially important when the objective is to compare scalar uncertainty metrics across candidate trajectories and to align the optimization objective with the resulting localization errors observed. Following the discussion in~\cite{ponda_trajectory_2009}, this choice is also motivated by the fact that, in azimuth-elevation localization problems, determinant-based \ac{FIM} criteria can become overly sensitive to range reduction and under-represent the benefit of angular separation between measurements.

Although the sensing geometry in this work is derived from a three-dimensional formulation using azimuth and elevation measurements, the case studies considered here reduce to a planar estimation problem. Specifically, the \acp{UAV} are constrained to operate at constant altitude and the surface vessels are assumed to operate in low sea-states, so the vertical component does not need to be estimated. As a result, the measurement model is three-dimensional in form, but the estimation problem is reduced to the horizontal coordinates. The rationale for adopting a trace-based metric therefore does not depend exclusively on solving a fully three-dimensional state-estimation problem. Rather, it remains appropriate here because the trace provides an interpretable measure of aggregate localization uncertainty in the reduced planar problem while retaining the geometric benefits identified in the broader three-dimensional setting.

In this work, the performance index follows the formulation of~\cite{ponda_trajectory_2009} by adopting the trace of the \ac{PCRLB}. The \ac{PCRLB}, introduced by Van Trees \emph{et al.}, provides a Bayesian lower bound on the estimator’s error covariance by incorporating prior uncertainty, information gained from new measurements, and process noise~\cite{VanTrees_2013}. In the Gaussian case, the \ac{PCRLB} defines an uncertainty ellipsoid around the estimated state, with the trace representing the sum of the variances along each principal axis, corresponding to the squared lengths of its semi-axes. Minimizing the trace of the \ac{PCRLB} over the planning horizon guides the UAV along trajectories that not only enhance measurement geometry, but also effectively reduce overall uncertainty given the prior knowledge. 

 \subsection{Numerical Optimization for Trajectory Design}
Due to the inherent nonlinearities and complex constraints in information-driven trajectory design, closed-form analytical solutions are generally intractable. Consequently, this work employs direct numerical optimization techniques that convert the infinite-dimensional optimal control problem into a finite-dimensional \ac{NLP} formulation, enabling the generation of dynamically feasible trajectories under realistic sensing, kinematic, and environmental constraints.

Although \ac{PS} techniques are commonly employed due to their fast convergence and high accuracy
~\cite{Robinson_2018}, the fundamental limitation of \ac{PS} methods is their non-uniform time discretization, which concentrates collocation points near the trajectory endpoints. Recent studies have demonstrated the utility of Bernstein polynomial basis functions over \ac{PS} methods~\cite{cichella_optimal_2018, cichella_optimal_2021}, as they offer several favorable advantages such as uniform time discretization and avoidance of the Runge phenomenon~\cite{FAROUKI_Bernstein}. Although they converge more slowly than \ac{PS} methods, these advantages alone make them an effective choice for use in trajectory generation where smooth state and control profiles are essential.

\subsection{Research Contributions}
This study advances estimation-aware trajectory optimization for bearing-only localization by incorporating mission constraints, dynamic constraints, and sensing constraints within a time-efficient optimization scheme. The formulation explicitly embeds \ac{FOV} limits, \ac{BTT} dynamics, \acp{NFZ}, and communication boundaries, enabling dynamically feasible and operationally compliant trajectories usable by low-\ac{SWaP-C} UAVs. In this initial study, both the adversarial target and the collaborating \ac{USV} are assumed stationary so that the effects of sensing geometry, coordination, and trajectory design can be isolated before extending the framework to moving-surface-vessel scenarios.

The conference version of this work focused on a single UAV~\cite{Mudrik_Sciacchitano_Kragelund_Kang_Kaminer_2025}, whereas this paper extends the work to coordinated multi-\ac{UAV} operations in three dimensions and demonstrates that optimized trajectories yield substantial improvements in estimator performance relative to heuristic paths. The \ac{PCRLB} defines a lower bound on the estimation error covariance matrix associated with the estimated state. The trace of this covariance matrix, which sums the variances along each state dimension, serves as a scalar measure of total position estimation uncertainty. The optimization results show that minimizing the trace of the \ac{PCRLB} yields more informative motion geometries and substantially reduces localization error in Monte Carlo validation.

The analysis further quantifies how \ac{EO} sensor selection, specifically cameras measuring visible light, influences achievable estimation accuracy. In particular, the results identify the minimum number of \ac{FFOV} \acp{UAV} with low-\ac{SWaP-C} required to match or exceed the localization performance of a single UAV equipped with a gimballed camera. Although time-varying surface-vessel motion is not treated in the present paper, the proposed formulation provides a baseline for future extensions to moving-target localization with more general kinematic models.

\subsection{Paper Organization}
The remainder of the paper is organized as follows. Section~\ref{Sec:Problem_Formulation} introduces the platform dynamics and sensor models, and formulates the joint trajectory optimization and estimation problem, incorporating key performance metrics derived from estimation theory.
Section~\ref{sec:case_studies} then details two illustrative case studies. The first evaluates optimized trajectories versus heuristic solutions for a single FFOV UAV, while the second compares the performance and \ac{SWaP-C} tradeoffs of a single UAV equipped with a gimballed sensor and a coordinated team of FFOV UAVs. Finally, Section~\ref{Sec:Conclusion} summarizes the research findings and highlights broader system-level insights derived from the study.

\section{Problem Formulation}\label{Sec:Problem_Formulation}
The trajectory optimization problem is defined for bearing-only localization using UAVs equipped with both gimballed and \ac{FFOV} sensors. In a \ac{GPS}-denied setting, the collaborating USV depends exclusively on the azimuth and elevation measurements from the UAVs to localize itself relative to the adversarial target. The objective is to design UAV flight paths that maximize localization accuracy by minimizing the trace of the PCRLB.

\subsection{Platform Kinematics}
Considering $N$ \acp{UAV}, each is initialized to operate at a low, constant altitude for the duration of the mission. This altitude constraint serves three key operational purposes: (1) it minimizes the risk of detection by adversarial sensors and weapon systems, (2) it reduces the potential for fratricide when operating in proximity to other UAVs, and (3) it enables a simplified dynamic model by eliminating vertical state dynamics. While three-dimensional localization is performed using azimuth and elevation information derived from bearing-only measurements~\cite{Ning_2024}, the fixed-altitude assumption justifies a reduction in the order of the UAV dynamics model. By constraining altitude, the vertical degree of freedom is removed, resulting in a planar (two-dimensional) dynamics formulation. This reduced-order model remains valid for surface-target engagement scenarios, where the elevation of stationary targets is effectively constant and does not influence maneuvering requirements.

To model the motion of UAV platforms, a \ac{BTT} configuration is adopted. The continuous-time equations of motion for the UAV under this configuration are
\begin{subequations}
	\begin{align}
		\dot{x}_{A}(t) & = V_{A} \cos \psi_{A}(t), \\
		\dot{y}_{A}(t) & = V_{A} \sin \psi_{A}(t), \\
		\dot{z}_{A}(t) & = 0, \\
		\dot{\psi}_{A}(t) & = \frac{g}{V_{A}} \tan \phi_{A}(t),
	\end{align}
	\label{eq:UAV_EOM}
\end{subequations}
where $x_{A}(t)$, $y_{A}(t)$, and $z_{A}(t)$ represent the UAV’s position in the inertial reference frame, $V_{A}$ is the constant airspeed, and $\psi_{A}(t)$ is the heading angle. The roll angle $\phi_{A}(t)$ serves as the control input that commands the turn rate $\dot{\psi}_{A}(t)$, which is governed by gravitational acceleration $g$ and the instantaneous roll command. Since excessive banking can lead to structural fatigue or instability, the roll angle is constrained by platform-specific limits

\begin{equation}
	\phi^{\min} \leq \phi_{A}(t) \leq \phi^{\max}, \quad \forall t \in [0,t_f],
\end{equation}

\noindent where $\phi^{\min}$ and $\phi^{\max}$ define the minimum and maximum allowable roll angles based on airframe design and maximum load factor. For each UAV indexed by $i$, where $i = 1, \ldots, N$, the state and control vectors for the platform are defined as
\begin{subequations}
    \begin{align}
        \textbf{x}_{A,i}(t) &\triangleq \begin{bmatrix}
		x_{A,i}(t) & y_{A,i}(t) & z_{A,i}(t) & \psi_{A,i}(t)
	\end{bmatrix}^\top, \\
        \textbf{u}_{A,i}(t) &\triangleq \phi_{A,i}(t),
    \end{align}
\end{subequations}
where the control input consists solely of the roll angle. 
This formulation applies identically to UAVs with both gimballed and body-fixed cameras.

For the \ac{UAV} equipped with a gimballed camera, additional degrees of freedom are introduced by enabling independent camera articulation in  azimuth and elevation, denoted as $\psi_g$ and $\theta_g$, respectively. This decouples the sensor pointing direction from the vehicle's heading, providing enhanced flexibility for target tracking. The problem formulation begins with a single UAV equipped with a gimballed camera, omitting the UAV index (subscript $i$) for notational clarity. 
Variables associated with the airframe are differentiated from variables associated with the camera gimbal by the subscripts $A$ and $g$, respectively, producing the extended state and control vectors

\begin{subequations}
    \begin{align}
        \mathbf{x}_{A,g}(t) &\triangleq \begin{bmatrix}
        x_A(t) & y_A(t) & z_A(t) & \psi_A(t) & \psi_g(t) & \theta_g(t)
        \end{bmatrix}^\top, \\
        \mathbf{u}_{A,g}(t) &\triangleq \begin{bmatrix}
        \phi_A(t) & u_{\psi_g}(t) & u_{\theta_g}(t)
        \end{bmatrix}^\top,
    \end{align}
\end{subequations}

\noindent where the gimbal angles are governed by single integrator dynamics
\begin{align}
\dot{\psi}_g(t) &= u_{\psi_g}(t), \\
\dot{\theta}_g(t) &= u_{\theta_g}(t).
\end{align}
The camera gimbal is constrained by angular rate limits to model the gimbal motor's actuators

\begin{align}
u_{\psi_g}^{\min} \leq u_{\psi_g}(t) \leq u_{\psi_g}^{\max}, \quad \forall t \in [0,t_f], \\
u_{\theta_g}^{\min} \leq u_{\theta_g}(t) \leq u_{\theta_g}^{\max}, \quad \forall t \in [0,t_f].
\end{align}

\begin{figure}[!htb]
    \centering
    \begin{subfigure}[b]{0.49\textwidth}
        \centering
        \includegraphics[width=\textwidth]{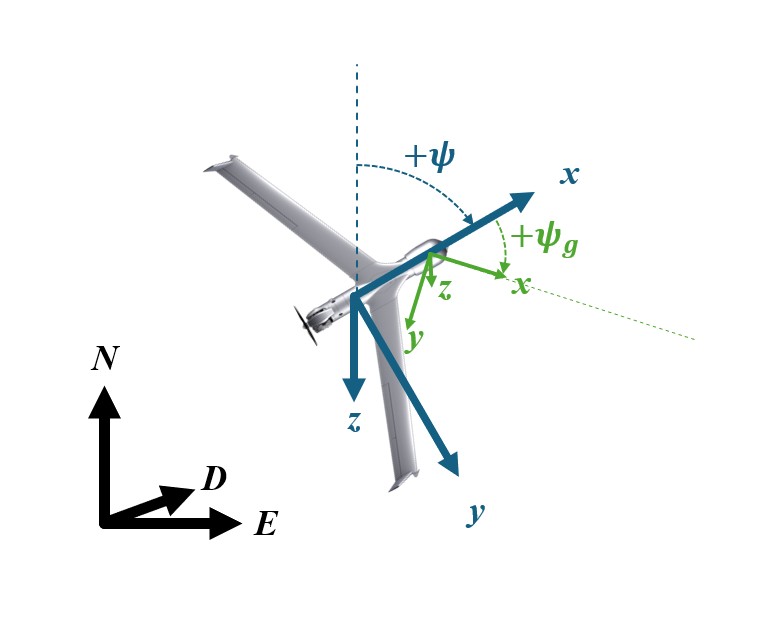} %
        \caption{Top-down view of gimbal reference frame.}
        \label{fig:Horizontal_Ref_Frame}
    \end{subfigure}
    \hfill
    \begin{subfigure}[b]{0.49\textwidth}
        \centering
        \includegraphics[width=\textwidth]{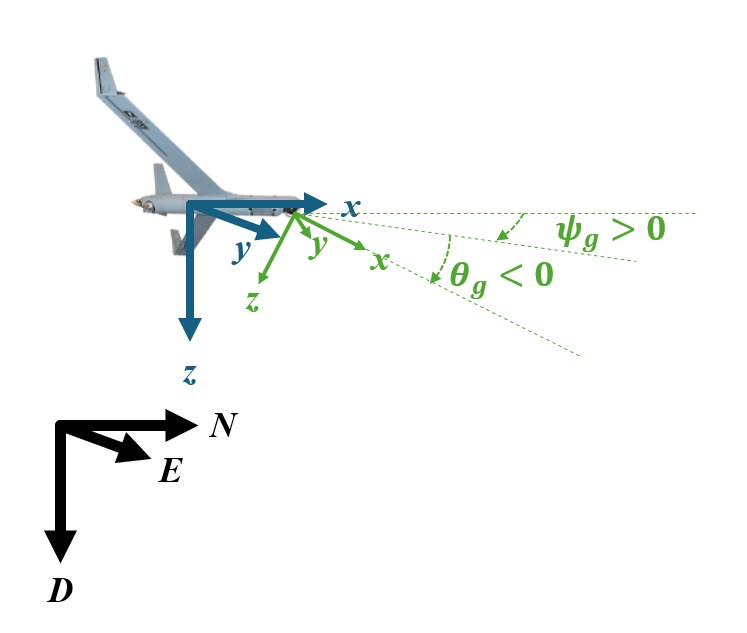} %
        \caption{Side view of gimbal reference frame.}
        \label{fig:Vertical_Ref_Frame}
    \end{subfigure}
    \caption{Reference frame depiction for a strap-down gimballed camera system relative to the UAV.}
    \label{fig:Gimbal_Ref_Frames}
\end{figure}

Figure~\ref{fig:Gimbal_Ref_Frames} illustrates the coordinate conventions for the gimballed camera frame from both top-down and side perspectives. The UAV body-fixed frame is shown in blue and follows the standard aerospace convention with axes $(x,y,z)$ pointing forward, right, and downward, respectively. The gimbal frame, depicted in green, introduces two additional rotational degrees of freedom. These are azimuth $\psi_g$, defined as rotation about the UAV's vertical
z-axis, and elevation $\theta_g$, defined as rotation about the gimbal’s local pitch axis. The Earth-fixed North-East-Down (NED) frame is shown in black for spatial reference.

\subsection{Operational Constraints}
To ensure effective target localization while preserving stealth, communication reliability, and overall mission feasibility, several operational constraints are imposed on UAV trajectories based on practical considerations such as minimizing the risk of counter-detection, enabling communications with the collaborating \ac{USV}, and ensuring valid sensor-target geometry. Specifically, three primary classes of constraints are considered: (1) no-fly zone enforcement around the adversarial target, (2) bounded communication range with the collaborating platform, and (3) overall mission area limits imposed by operational boundaries.

\subsubsection{No-Fly Zone Around the Target}

To reduce the likelihood of counter-detection by adversarial systems, an \ac{NFZ} is defined around the target of interest. \ac{UAV} operations are constrained to remain outside this zone to avoid detection, classification, or engagement by the target's systems. The radius of the \ac{NFZ} is selected based on the target's detection capabilities, the UAV’s stealth profile, and environmental risk factors. The position vectors, $\textbf{p}_{A,i}$, for each UAV indexed by $i$, are

\begin{equation}
    \mathbf{p}_{A,i}(t) = %
    \begin{bmatrix}
        x_{A,i}(t) & y_{A,i}(t) & z_{A,i}(t)
    \end{bmatrix}^\top
\end{equation}
The position of the target USV, $\textbf{p}_{T}$, is

\begin{equation}
    \mathbf{p}_{T}(t) = 
    \begin{bmatrix}
        x_{T}(t) & y_{T}(t) & z_{T}(t)
    \end{bmatrix}^\top.
\end{equation}  
Let $r_{\text{NFZ}}$ denote the radius of the \ac{NFZ}. The constraint is imposed as

\begin{equation}
    \|\mathbf{p}_{T}(t)- \mathbf{p}_{A,i}(t)  \| \geq r_{\text{NFZ}}, \quad \forall t\in [0,t_f]
\end{equation}
where $\| \cdot \|$ denotes the Euclidean norm. This geometric exclusion ensures that UAVs remain outside the target’s effective detection range throughout the mission timeline.

\subsubsection{Maximum Communication Range}
 The \acp{UAV} must establish a communication link with the collaborating USV to exchange critical targeting data upon the completion of target localization. To maintain effective communication while ensuring security, a boundary condition is imposed on the maximum communication range at the terminal time, $t_{f}$. This constraint ensures that collected data can be transmitted effectively to the cooperative platform without risking counter-detection by external threats or increasing individual platform \ac{SWaP-C} by requiring high-power communication systems for transmission. 
 
Let $r_{\text{com}}$ denote the maximum allowable communication range, and let
\begin{equation}
    \mathbf{p}_{USV}(t) = 
    \begin{bmatrix}
        x_{USV}(t) & y_{USV}(t) & z_{USV}(t)
    \end{bmatrix}^\top
\end{equation}
be the position of the collaborating USV. The constraint is expressed as

\begin{equation}
    \| \mathbf{p}_{USV}(t_{f}) - \mathbf{p}_{A,i}(t_{f}) \| \leq r_{\text{com}}
\end{equation}

This communication boundary plays a critical role in preserving the functional integrity of the coordination strategy by enforcing a maximum boundary in which electromagnetic emissions are allowed.

\subsubsection{Mission Area Bounds}

In addition to local constraints near the target and collaborating USV, the UAV must operate within a bounded mission area that reflects the limits of available airspace, sensor coverage, and operational risk tolerance. These bounds are not imposed based on globally persistent knowledge or absolute geolocation, but are derived from prior mission planning, relative sensor constraints, and mission-relevant regions of interest.

Let the mission area be defined by a rectangular region aligned with the inertial coordinate axes. The admissible operating domain for each UAV, $i$, is bounded by
\begin{align}
&x_{\min} \leq x_{A,i}(t) \leq x_{\max}, \quad \ \forall t\in [0,t_f] \\
&y_{\min} \leq y_{A,i}(t) \leq y_{\max}, \quad \forall t\in [0,t_f]
\end{align}
  
\noindent where $x_{\min}, x_{\max}, y_{\min}$, and $y_{\max}$, define the spatial extents of the mission area in the horizontal plane. Enforcing these bounds constrains UAV motion to mission-relevant regions, improves numerical conditioning of the optimization by avoiding infeasible or unbounded trajectories, and enables local enforcement using onboard sensors or cooperative data sharing. Collectively, the \ac{NFZ}, communication range, and mission area bounds define the spatial limits wherein each UAV must operate. These constraints are explicitly incorporated into the trajectory optimization formulation to ensure all generated trajectories remain feasible and mission compliant.

\subsection{Sensor Modeling}
Accurate modeling of the \ac{UAV}’s onboard camera sensor is essential for enabling realistic trajectory optimization with bearing only measurements. The information gained from sensor measurements depends on the relative geometry between the UAV and the target(s), as well as sensor \ac{FOV} constraints, which determine whether a given measurement is valid and how effectively it contributes to reducing target localization error. Here, information refers to the degree to which a measurement constrains the estimated position of the target, based on sensor-target geometry.

This section details the coordinate transformations required to compute relative measurements in the sensor frame, models the \ac{FOV} constraint using smooth sigmoid-based functions to ensure compatibility with gradient-based optimization, and incorporates the augmented dynamic model needed for a UAV equipped with a gimballed camera. While the physical \ac{FOV} is defined by strict angular limits, these are approximated using differentiable sigmoid functions to enable efficient numerical optimization~\cite{Kragelund_Walton_Kaminer_Dobrokhodov_2021}. The approximation captures the essential geometric constraints of the actual sensor model while avoiding the discontinuities associated with hard-threshold \ac{FOV} enforcement.

\subsubsection{Relative Position and Frame Transforms}
Let $^{I}\mathbf{p}_{c,i}(t) \triangleq [x_{A,i}(t),\, y_{A,i}(t),\, z_{A,i}(t)]^{\top}$ represent the camera position in a common local \ac{NED} inertial frame for UAV $i$, assuming the camera is collocated with the UAV’s center of mass. In this work, the index $j$ is used generically to denote either the adversarial target USV or the collaborating USV, both of which are expressed in the same inertial mission frame. Their inertial positions are therefore written as $^{I}\mathbf{p}_{j}(t) \triangleq [x_{j}(t),\, y_{j}(t),\, z_{j}(t)]^{\top}$. %
Position relative to the UAV is 
\begin{equation}
    ^{I}\Delta\mathbf{p}_{i,j}(t)\;\triangleq\;^{I}\mathbf{p}_{j}(t)\;-\;^{I}\mathbf{p}_{c,i}(t).\;%
\end{equation}

\begin{figure}[htbp]
    \centering
    \includegraphics[width=0.5\linewidth]{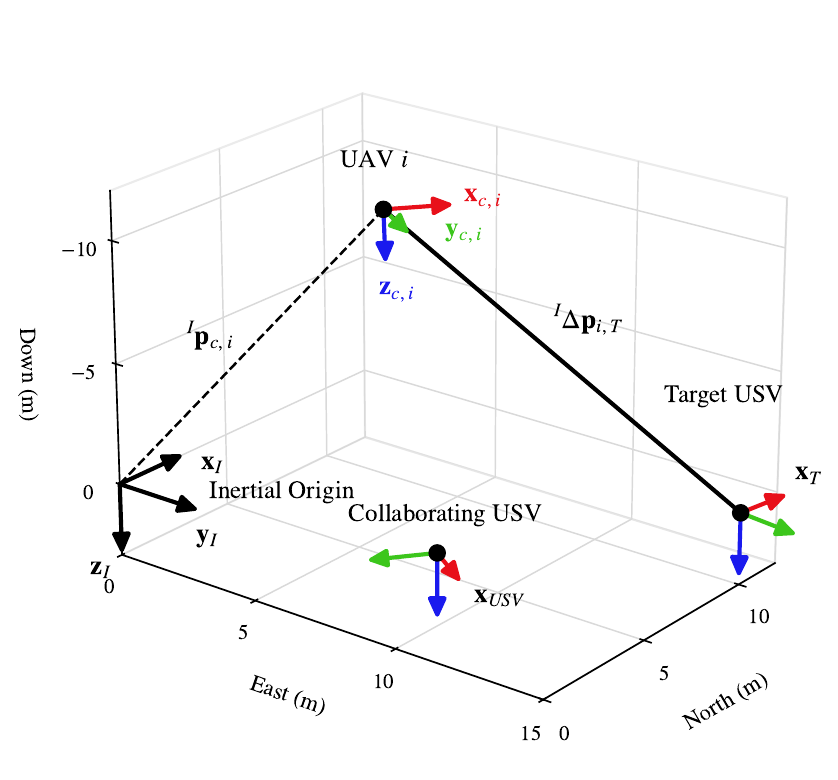}
    \caption{Reference-frame geometry for the measurement model. UAV and surface-vessel positions are expressed in a local inertial frame, then transformed to the camera frame for bearing computation.}
    \label{fig:Inertial Frame}
\end{figure}

\noindent Position with respect to the UAV’s body-fixed camera frame is

\begin{equation}
   ^{c}\mathbf{p}_{i,j}(t)\;=\; \mathbf{R}_{I}^{c,i}(t)\;^{I}\Delta\mathbf{p}_{i,j}(t)
\end{equation}

\noindent where $\textbf{R}_{I}^{c,i}(t)$ denotes the rotation matrix from the inertial to the camera frame of UAV $i$, constructed from sequential rotations by the yaw, pitch, and roll angles, respectively
\begin{equation}
\mathbf{R}_{I}^{c,i}(t) = \mathbf{R}\bigl(\phi_{i}(t)\bigr)\, \mathbf{R}\bigl(\theta_{i}(t)\bigr)\, \mathbf{R}\bigl(\psi_{i}(t)\bigr).
\end{equation}
Figure~\ref{fig:Inertial Frame} illustrates the relationship between the common inertial mission frame, the UAV camera frame, and the positions of both the adversarial target USV and the collaborating USV. This additional diagram is included to clarify the full reference-frame structure used throughout the measurement model.

For a single UAV equipped with a gimballed camera, we can omit the subscript $i$ and express this relative position in the gimbal frame with two additional rotations by the commanded azimuth $\psi_g$ and elevation $\theta_g$ angles

\begin{equation}
  ^{c_g}\mathbf{p}_{j}(t)\;=\; \mathbf{R}_{c}^{c_g}(t)\;^{c}\mathbf{p}_{j}(t),
\end{equation}
where
\begin{equation}
    \mathbf{R}_{c}^{c_g}(t) = \mathbf{R}\bigl(\theta_{g}(t)\bigr)\, \mathbf{R}\bigl(\psi_{g}(t)\bigr).
\end{equation}

For either camera type, the sensor extracts azimuth ($\alpha_j$) and elevation ($\beta_j$) bearing angles as

\begin{subequations}
    \begin{align}
        \alpha_{i,j}(t) &\triangleq \mathrm{atan2}\!\left(^{c,i}r_{y,i,j}(t),\,^{c,i}r_{x,i,j}(t)\right), \\
        \beta_{i,j}(t)  &\triangleq \tan^{-1}\!\left(\frac{^{c,i}r_{z,i,j}(t)}
        {\sqrt{(^{c,i}r_{x,i,j}(t))^2 + (^{c,i}r_{y,i,j}(t))^2}}\right).
    \end{align}
\end{subequations}

The azimuth angle is written using the two-argument inverse tangent to preserve quadrant information, whereas the elevation angle is written in its conventional form as the angle between the LOS vector and its horizontal projection. This distinction preserves a unique geometric interpretation for each bearing component.
\subsubsection{FOV Constraints}

Each camera has a limited \ac{FOV} defined by azimuth and elevation bounds. The target and the collaborating USV, indexed by $j$, are considered visible to UAV $i$ if their camera-frame azimuth and elevation angles satisfy the following angular \ac{FOV} constraints

\begin{equation}
\alpha_{\min} \leq \alpha_{i,j}(t) \leq \alpha_{\max}, \quad
\beta_{\min} \leq \beta_{i,j}(t) \leq \beta_{\max}.
\label{eq: FOV limits}
\end{equation}

When either angle exceeds these limits, an out-of-frame event occurs, meaning the target temporarily leaves the sensor’s \ac{FOV}. These events interrupt the acquisition bearing measurements and reduce information gain. To account for this effect in the optimization, visibility is modeled as a continuous weighting function that smoothly decreases to zero as the target approaches the FOV boundary. This allows the optimizer to anticipate and minimize information loss due to these out-of-frame conditions.

To enable gradient-based optimization over real-valued decision variables, each angular visibility constraint in Eq.~\eqref{eq: FOV limits} is approximated using a smooth, parameterized sigmoid function. This produces a continuously differentiable surrogate for the original discontinuous indicator function, which outputs values in $[0,1]$ and approximates binary visibility decisions in a differentiable manner. The sigmoid function for each angular constraint is

\begin{equation}
\mathbb{I}_{\delta_{\mathrm{limit}}}= \frac{1}{1 + e^{k(\delta - \delta_{\mathrm{limit}})}},
\label{eq: Indicator fxn}
\end{equation}

\noindent where $\delta$ is the measured angle, $\delta_{\mathrm{limit}}$ is the threshold, and $k > 0$ controls the steepness of the transition. 

\begin{figure}[htb!]
    \centering
    \includegraphics[width=0.7\textwidth]{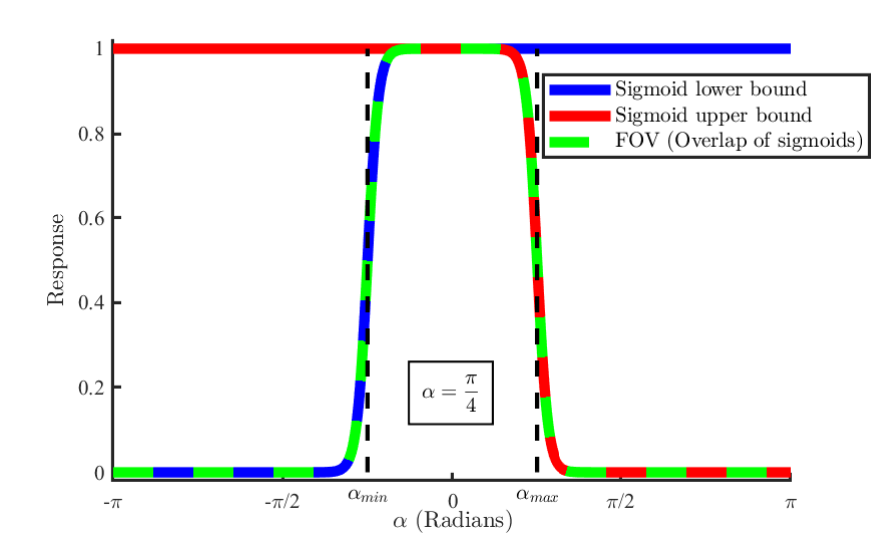}
    \caption{Smooth approximation of FOV constraints using overlapping sigmoid functions.}
    \label{fig:SigmoidExample}
\end{figure}

The complete approximation of the FOV indicator function, denoted $\mathbb{I}_{FOV_{i,j}}(t)$, is then constructed as the product of four sigmoid-based functions that correspond to the upper and lower bounds on the azimuth and elevation angles in the sensor FOV,

\begin{equation}
\mathbb{I}_{FOV_{i,j}}(t) =
\mathbb{I}_{\alpha_{\min},i,j}(t) \cdot 
\mathbb{I}_{\alpha_{\max},i,j}(t) \cdot 
\mathbb{I}_{\beta_{\min},i,j}(t) \cdot 
\mathbb{I}_{\beta_{\max},i,j}(t),
\label{eq:FOV_indicator_product}
\end{equation}

\noindent where each term smoothly enforces one side of the angular visibility, ensuring that $\mathbb{I}_{FOV_{i,j}}(t) \in [0,1]$. Fig.~\ref{fig:SigmoidExample} illustrates how overlapping sigmoid functions approximate the angular visibility bounds for azimuth angles $\alpha \in [-\frac{\pi}{4},\frac{\pi}{4}]$.

The resulting visibility function $\mathbb{I}_{FOV_{i,j}}(t)$, provides a continuous and differentiable approximation of target visibility. This property preserves the smoothness required for use with gradient-enabled numerical solvers~\cite{Casadi_2019}, and ensures that information weighting remains physically meaningful during out-of-frame transitions.

\subsection{Performance Index -- The PCRLB}

To quantify the estimation performance of a candidate UAV trajectory, the \ac{PCRLB} is adopted as the primary performance index. The \ac{PCRLB} provides a theoretical lower bound on the covariance of any Bayesian estimator and serves as a proxy for assessing localization accuracy in the bearing-only tracking scenarios considered.  

The inclusion of the prior is critical in this work because both the target and the collaborating USV are assumed stationary. The prior represents the initial covariance of the state estimate before any bearing measurements are obtained, reflecting the uncertainty associated with pre-mission cueing or previous sensing. In practice, this prior may originate from external intelligence or surveillance assets such as radar, satellite imagery, or surface-based sensors that provide coarse initial localization of the target and collaborating USV. As new measurements are collected, this prior information is updated through the estimation process, yielding a posterior bound that reflects the cumulative information gained. 

The trace of this matrix, which sums the variances along each state dimension, serves as a scalar measure of total position estimation uncertainty. This scalar quantity is conceptually analogous to the \ac{PDOP} used in GPS-based systems, where PDOP quantifies how the geometric configuration of satellites amplifies the impact of measurement noise on position accuracy. Similarly, minimizing the trace of the PCRLB favors trajectories that improve sensing geometry and reduce the theoretical lower bound on estimation error, thereby enhancing overall localization performance. 

\ac{EO} sensors in general operate at fixed sampling frequencies, therefore, this work adopts a discrete-time approach for modeling bearing measurements. This formulation aligns naturally with how sensor data is generated and processed onboard the \ac{UAV}, as measurements arrive at regular intervals determined by hardware and computational constraints. Under these conditions, the information relevant to estimation performance also accumulates discretely in time, making the discrete formulation of the \ac{FIM} and \ac{PCRLB} both practical and consistent with standard filtering and numerical optimization methods.

The recursive form of the PCRLB used in this work follows the formulation of~\cite{tichavsky_posterior_1998}, where the prior information is propagated through the \ac{FIM} to account for both prediction and measurement updates. For the stationary target case considered here, the recursion simplifies to a sequential accumulation of bearing measurement information, starting from the initial covariance corresponding to the prior uncertainty. This ensures that the resulting bound reflects both the initial knowledge of the target state and the incremental information gained from new measurements. The posterior bound satisfies

\begin{equation}
\mathbb{E}\{[\hat{\mathbf{x}}_{k|k} - \mathbf{x}_{k}][\hat{\mathbf{x}}_{k|k} - \mathbf{x}_{k}]^\top\} \geq \mathbf{PCRLB}(k) = \mathbf{FIM}(k)^{-1},
\end{equation}

\noindent where $\hat{\mathbf{x}}_{k|k}$ is the estimated target state and $\mathbf{x}_k$ is the true state at time $k$.

The \ac{FIM} for a stationary target $j$, observed by UAV $i$ over $\text{N}_{\text{k}}$ discrete time steps, is computed according to standard formulations in estimation theory~\cite{VanTrees_2013, ristic_beyond_2003} as

\begin{equation}\label{eq:FIM_discrete_3D}
\mathbf{FIM}_j = 
\sum_{k=1}^{\text{N}_{\text{k}}}  \left( \frac{\partial \boldsymbol{h}_{i,j}(\mathbf{x}_i(k),\mathbf{x}_j(k))}{\partial \mathbf{x}_j(k)} \right)^\top \mathbf{R}_{i,j}^{-1}(k) \left( \frac{\partial \boldsymbol{h}_{i,j}(\mathbf{x}_i(k),\mathbf{x}_j(k))}{\partial \mathbf{x}_j(k)} \right),
\end{equation}

\noindent where
$\boldsymbol{h}_{i,j}(\mathbf{x}_i(k),\mathbf{x}_j(k)) = [\alpha_{i,j}(k), \beta_{i,j}(k)]^\top$ contains the azimuth and elevation angles from UAV $i$ to target $j$ and $\mathbf{R}_{i,j}(k)$ is the corresponding measurement noise covariance matrix

\begin{equation}
\mathbf{R}_{i,j}(k) =
\begin{bmatrix}
\sigma_{\alpha_{i,j}}^2(k) & 0 \\
0 & \sigma_{\beta_{i,j}}^2(k)
\end{bmatrix}.
\end{equation}

The measurement functions for azimuth and elevation are defined in an NED coordinate frame by the geometric relations

\begin{equation}
    \boldsymbol{h}_{i,j}(\mathbf{x}_i(k),\mathbf{x}_j(k)) \triangleq 
        \begin{bmatrix}
        {\alpha_{i,j}} \\
        {\beta_{i,j}} 
    \end{bmatrix},
\end{equation}

\noindent where $\mathbf{x}_i(k)$ are the known positions of UAV $i$, assumed to be available from prior information such as third-party targeting or mission planning data and maintained via onboard inertial measurements, and $\mathbf{x}_j(k)$ are the unknown target states to be estimated.

The Jacobian of the measurement function $\boldsymbol{h}_{i,j}(\mathbf{x}_i(k),\mathbf{x}_j(k))$ with respect to the target state $\mathbf{x}_j(k)$ is adapted from \cite{ponda_trajectory_2009} and is given by

\begin{equation}
    \frac{\partial \boldsymbol{h}_{i,j}(\mathbf{x}_i(k),\mathbf{x}_j(k))}{\partial \mathbf{x}_j(k)} =
    \begin{bmatrix}
        \frac{\partial \alpha_{i,j}}{\partial x_j} & \frac{\partial \alpha_{i,j}}{\partial y_j} \\ %
        \frac{\partial \beta_{i,j}}{\partial x_j}  & \frac{\partial \beta_{i,j}}{\partial y_j} \\ %
    \end{bmatrix}_k
    =
        \begin{bmatrix}
        -\frac{r_{y,i,j}}{r_{x,i,j}^2 + r_{y,i,j}^2} & \frac{r_{x,i,j}}{r_{x,i,j}^2 + r_{y,i,j}^2}\\ %
        -\frac{r_{x,i,j}r_{z,i,j}}{r_{i,j}^2 \sqrt{r_{x,i,j}^2 + r_{y,i,j}^2}}  & -\frac{r_{y,i,j}r_{z,i,j}}{r_{i,j}^2 \sqrt{r_{x,i,j}^2 + r_{y,i,j}^2}}  \\ %
    \end{bmatrix}_k,
\end{equation}
where $r_{i,j}$ is the euclidean distance between \ac{UAV} $i$ and target $j$. Note that although the target and UAV positions include a z-component, changes in the vertical direction are not estimated. Since the UAV and target altitudes are assumed to be known and constant, the measurement Jacobian is reduced to a $2\times2$ form involving only the x and y components of the target state.  

To reflect realistic sensing constraints, each FIM contribution is weighted by the continuous visibility factor $\mathbb{I}_{FOV_{i,j}}(k)$ defined in Eq.~\eqref{eq:FOV_indicator_product}. Additionally, external cuing is incorporated in the initialization of the FIM where the initial prior enters the information matrix as $\mathbf{P}_{0,j}^{-1}$. This prior is distributed across all UAVs using a uniform weighting rule 

\begin{equation}
    \lambda_i = \frac{1}{N_{\mathrm{UAV}}}, \qquad \sum_{i=1}^{N_{\mathrm{UAV}}}\lambda_i = 1,
    \label{eq:weight_factor}
\end{equation}

\noindent where $\lambda_i$ are the uniform weights that sum to one. The fused formulation accounting for the combined information gain by all UAVs is given by

\begin{equation}\label{eq:FIM_local_FOV}
\mathbf{FIM}_{j} = \sum_{i=1}^{N_{\mathrm{UAV}}}\biggl\{\lambda_i\,\mathbf{P}_{0,j}^{-1}
    + \sum_{k=1}^{N_k} 
    \mathbb{I}_{FOV_{i,j}}(k) \left( \frac{\partial \boldsymbol{h}_{i,j}(\mathbf{x}_i(k),\mathbf{x}_j(k))}{\partial \mathbf{x}_j(k)} \right)^{\!\top}
    \mathbf{R}_{i,j}^{-1}(k)
    \left( \frac{\partial \boldsymbol{h}_{i,j}(\mathbf{x}_i(k),\mathbf{x}_j(k))}{\partial \mathbf{x}_j(k)} \right) \biggr\}  
\end{equation}

The resulting PCRLB, derived from the inverse of this FIM, represents a lower bound on the estimation error covariance  for target $j$. Minimizing the trace of the \ac{PCRLB} was found to be an effective performance index for optimization in \cite{ponda_trajectory_2009}. Therefore, we utilize the trace of the \ac{PCRLB} in this work to facilitate direct performance comparisons across different UAV configurations and mission scenarios.

\subsection{Optimization}\label{Sec: Optimization}
The trajectory design problem is formulated as a constrained optimization problem that minimizes a weighted objective function combining estimation performance and mission duration. This multi-objective formulation captures the trade-off between target localization performance and time efficiency, subject to the kinematic, sensing, and operational constraints introduced in prior sections.

The cost function incorporates the trace of the \ac{PCRLB} for both the target, $\text{PCRLB}_{\text{T}}$, and the collaborating USV, $\text{PCRLB}_{\text{USV}}$. Tunable parameters, $w_1, w_2 \in [0,1]$, are used to control their relative importance for localization, respectively. An additional penalty term scaled by $w_3 \geq 0$ is included to penalize unnecessarily long trajectories. The cost function formulation for $N$ UAVs is

\begin{equation}\label{eq:cost_function}
    \min_{\mathbf{x}(t),\mathbf{u}(t)} \quad
    J \;=\;  w_1 \cdot \mathrm{Tr}\,\!\bigl(\mathrm{PCRLB}_{T}\bigr)
    + w_2 \cdot \mathrm{Tr}\,\!\bigl(\mathrm{PCRLB}_{\text{USV}}\bigr)
    + w_3 \cdot t_{f},
\end{equation}

\noindent where $\mathbf{x}(t)$ and $\mathbf{u}(t)$ denote the state and control trajectories for all UAVs, $\mathrm{Tr}(\cdot)$ is the matrix trace operator, and $t_f$ is the final time of the mission. This cost function is subject to the following constraints
\begin{subequations}
\begin{align}
    &\dot{\mathbf{x}}_{i}(t) = f(\mathbf{x}_{i}(t), \mathbf{u}_{i}(t)), \\
    &\phi^{\min} \leq \phi_{A}(t) \leq \phi^{\max}, \quad &\forall t \in [0, t_f], \\
        &x_{\min} \leq x_{A,i}(t) \leq x_{\max}, \quad &\forall t \in [0, t_f], \\
&y_{\min} \leq y_{A,i}(t) \leq y_{\max}, \quad &\forall t \in [0, t_f],\\
    &\|\mathbf{p}_{T}(t) - \mathbf{p}_{A,i}(t)\| \geq r_{\text{NFZ}}, \quad &\forall t \in [0, t_f], \\
    &\|\mathbf{p}_{\text{USV}}(t_f) - \mathbf{p}_{A,i}(t_f)\| \leq r_{\text{com}}.
\end{align}
\end{subequations}

These constraints ensure that the resultant UAV trajectories are dynamically feasible and do not violate operational considerations. However, because the trajectory and control profiles are continuous functions of time, this formulation represents an infinite-dimensional optimization problem whose exact solution is generally intractable. This motivates the use of numerical approximation methods to obtain computationally feasible solutions.

\subsection{Bernstein Approximation}
To obtain a tractable formulation, the problem is transformed into a finite-dimensional \ac{NLP} problem by approximating the continuous-time state and control trajectories for each UAV using Bernstein polynomials. This parameterization is particularly advantageous because the Bernstein basis is the only known polynomial basis that operates with uniformly spaced time nodes, ensuring compatibility with fixed sensor sampling rates and avoiding the Runge phenomenon observed in other polynomial approximations~\cite{FAROUKI_Bernstein}. 

The Bernstein basis polynomial of degree $n$ is defined as

\begin{equation}
	b_{j,n}(t) = \binom{n}{j} \frac{t^{j} (t_{f}-t)^{n-j}}{t^{n}_{f}}, \quad  j = 0, \dots,n, \quad\forall t \in [0,t_{f}]
\end{equation}

\noindent where the binomial coefficient is

\begin{equation}
	\binom{n}{j} = \frac{n!}{j!(n-j)!}
\end{equation}

\noindent The corresponding time nodes are equidistant and given by

\begin{equation}
	t_j = \frac{j}{n}t_f, \quad j = 0, \dots, n
\end{equation}

\noindent where $t_{f} > 0$ is given. The state and control trajectories are then approximated as linear combinations of Bernstein basis functions
\begin{align}
	\mathbf{x}(t) &\approx \sum_{j=0}^{n} \bar{\mathbf{x}}_{j,n} b_{j,n}(t), \\
	\mathbf{u}(t) &\approx \sum_{j=0}^{n} \bar{\mathbf{u}}_{j,n} b_{j,n}(t),
\end{align}

\noindent where $\bar{\mathbf{x}}_{j,n} \in \mathbb{R}^{n_x}$ and $\bar{\mathbf{u}}_{j,n} \in \mathbb{R}^{n_u}$ are the Bernstein coefficients for the states and controls, respectively. These coefficients are used to define the matrices
\begin{align}
	\bar{\mathbf{x}}_n &= \begin{bmatrix} \bar{\mathbf{x}}_{0,n} & \cdots & \bar{\mathbf{x}}_{n,n} \end{bmatrix} \in \mathbb{R}^{n_x \times (n+1)}, \\
	\bar{\mathbf{u}}_n &= \begin{bmatrix} \bar{\mathbf{u}}_{0,n} & \cdots & \bar{\mathbf{u}}_{n,n} \end{bmatrix} \in \mathbb{R}^{n_u \times (n+1)}
\end{align}

A key advantage of Bernstein polynomials is their well-defined differentiation property, which applies to any function approximated by the Bernstein basis 
	\begin{align}
		\dot{\textbf{x}}_{n}(t) & = \sum_{j=0}^{n-1} \frac{n}{t_{f}} (\bar{\textbf{x}}_{j+1,n} - \bar{\textbf{x}}_{j,n}) b_{j,n-1}(t) %
		 = \sum_{j=0}^{n-1} \sum_{i=0}^{n} \bar{\textbf{x}}_{i,n} \textbf{D}_{i,j}  b_{j,n-1}(t) 
		\label{eq:dyn_ber}
	\end{align}
\noindent where $\textbf{D}_{i,j} $ is the $(i,j)$-th entry of the differentiation matrix as described in \cite{cichella_optimal_2021}.

For a dynamic model satisfying $ \dot{\mathbf{x}}_{i}(t) = f(\mathbf{x}_{i}(t), \mathbf{u}_{i}(t))$ and a number $1 > \delta > 0$, this differential constraint is approximated using a relaxation 
	\begin{equation}
		\left\Vert \sum_{i=0}^{n} \bar{\textbf{x}}_{i,n} \textbf{D}_{i,j} - \textbf{f}(\bar{\textbf{x}}_{j,n},\bar{\textbf{u}}_{j,n}) \right\Vert \leq \delta^{n}, \qquad \forall j = 0, \dots, n
		\label{eq:convergence}
	\end{equation}
\noindent where $\delta^{n}$ is a relaxation that converges to zero as $n \to \infty$. Another important property is that any Bernstein approximation lies within the convex hull of its control points

	\begin{equation}
		\min_{j=0,\dots,n} \bar{\textbf{x}}_{j,n} \leq \: \textbf{x}_{n}(t) \: \leq \max_{j=0,\dots,n} \bar{\textbf{x}}_{j,n} \qquad \forall t \in [0,t_{f}]
		\label{eq:Ch}
	\end{equation}

This property allows inequality constraints, such as the mission boundaries, to be enforced directly on the Bernstein control points, $\bar{\textbf{x}}_{j,n}$.  Since the trajectory lies within the convex hull of its control points, %
enforcing these constraints at the control points guarantees they are satisfied over the entire continuous trajectory, which simplifies constraint enforcement during the optimization process. For nonconvex geometric constraints such as exclusion regions, the convex-hull property alone does not in general provide a global guarantee. In particular, when the UAV encircles the target, the \ac{NFZ} may lie inside the convex hull of the control points even though the continuous trajectory itself remains feasible. In this case, the convex-hull argument is still useful for conservative reasoning, but it should not be interpreted as a universal guarantee for exclusion-type constraints.
    
Further details on the use of Bernstein polynomials for trajectory generation problems are provided in~\cite{FAROUKI_Bernstein,Mehdi_Choe_Cichella_Hovakimyan_2015, cichella_optimal_2018,cichella_optimal_2021,kielas-jensen_bernstein_2022}

\subsection{Federated Sensor Fusion via Extended Kalman Filtering}
One architecture for multi-UAV target localization is a centralized approach in which the collaborating \ac{USV} receives all bearing-only measurements from the UAVs and processes them within a single, global \ac{EKF}. While this method yields consistent and fully fused estimates, it requires the collaborating USV to maintain continuous, high-bandwidth communication with all UAVs throughout the mission, which incurs a higher \ac{SWaP-C} burden.

To overcome this limitation, this work employs a \ac{FKF}  that conducts data fusion at the conclusion of the mission. For coordinated target localization by multiple FFOV UAVs, each UAV independently tracks the target and collaborating \acp{USV} using a local \ac{EKF} based on its own bearing measurements. These local EKFs run in parallel across the team of UAVs throughout the mission, producing time-varying state estimates $\mathbf{x}_{i,j}(k)$ and associated covariance matrices $\mathbf{P}_{i,j}(k)$ for each target $j$.

To synthesize these estimates into a single, system-level estimate, an \ac{FKF} is applied at the final time step, after each UAV has completed its data collection and returned to the collaborating USV. This federated fusion step enables centralized aggregation of all local estimates using an information-weighted consensus rule. The \ac{FKF} is computed according to standard estimation formulations \cite{carlson_federated_1990,carlson_federated_1994,Kim_Jee_Lee_1998} as

\begin{equation}
    \mathbf{x}_{j}(k_f) = \left( \sum_{i=1}^{\text{N}_{\text{UAV}}} \mathbf{P}_{i,j}(k_f)^{-1} \right)^{-1} \sum_{i=1}^{\text{N}_{\text{UAV}}} \mathbf{P}_{i,j}(k_f)^{-1} \mathbf{x}_{i,j}(k_f),
\end{equation}
\begin{equation}
    \mathbf{P}_{f}(k_f) = \left( \sum_{i=1}^{\text{N}_{\text{UAV}}} \mathbf{P}_{i,j}(k_f)^{-1} \right)^{-1},
\end{equation}

\noindent where $k_f$ denotes the final fusion time, and $\text{N}_{\text{UAV}}$ is the number of participating UAVs. This formulation ensures consistent fusion without requiring inter-UAV communication or centralized processing during flight, making it compatible with bandwidth-limited systems such as those designed for low-\ac{SWaP-C}.

The initialization of each local EKF is tied to the trajectory optimization process. Prior to deployment, each UAV is initialized with a common estimate of the target and USV states, derived from third-party cuing or prior mission knowledge.  This shared prior defines the initial state and covariance used for estimator initialization. To prevent double counting during federated fusion, the initialization of the prior information is partitioned across UAVs as shown in Eq.~\eqref{eq:weight_factor}. Importantly, this is the same covariance used to initialize the trajectory optimization, ensuring that planned trajectories are shaped to reduce uncertainty relative to the actual estimator initialization. As a result, both planning and estimation are aligned to maximize the value of each UAV’s individual measurement given the expected state uncertainty at mission start.

By concentrating data fusion at the terminal step, this FKF-based approach enables robust, low-bandwidth cooperative localization without sacrificing estimator consistency. It supports scalable, federated architectures where FFOV UAVs can operate independently yet still contribute effectively to a shared targeting solution in GPS-denied or electromagnetically contested environments.

\section{Results}
\label{sec:case_studies}

Traditionally, UAV path planning for localization has relied on heuristic trajectories. These trajectories, such as the racetrack pattern or circular orbit, are easy to implement and are consistent with a legacy doctrine. While operationally convenient, these paths may fail to maximize information gain. In contrast, trajectories optimized with an information-theoretic objective such as minimizing the trace of the \ac{PCRLB}, can significantly enhance localization accuracy and estimator performance. By explicitly accounting for time-varying geometry, visibility constraints, and sensor \ac{FOV} limitations, these optimized paths are tailored to mission objectives, thereby enabling more intelligent and effective UAV operations.

Beyond performance metrics, trajectory optimization must be considered not only from the perspective of estimation performance but also within the broader context of system-level constraints. \ac{SWaP-C} considerations, especially those tied to the airframe and sensor configuration, have a direct impact on platform design and operational deployment. Optimization may reduce flight time or the number of vehicles required, but platform-level design tradeoffs, such as the use of mechanically gimballed versus \ac{FFOV} camera sensors, also play a pivotal role in determining system suitability.  These tradeoffs enable trajectory optimization to emerge as both a control-theoretic enhancement and an enabler of more affordable and deployable system designs.

The interdependence of estimation accuracy, flight path design, and hardware configuration is highlighted in the case studies that follow. The first case study compares a heuristic racetrack trajectory with an optimized path for a single FFOV UAV, focusing on how trajectory shape alone influences localization performance.

The second case study evaluates the tradeoffs between a single UAV equipped with a gimballed camera and a coordinated team of FFOV \acp{UAV}. While the gimballed platform benefits from independent sensor articulation, it incurs a higher \ac{SWaP-C} burden. The FFOV team, however, leverages multi-UAV coordination to match or exceed localization performance, demonstrating that distributed sensing systems can achieve high fidelity at reduced cost and greater redundancy. 

Throughout the results, the scalar bound $\sqrt{\mathrm{Tr}(\mathrm{PCRLB})}$ is used to enable a direct, unit-consistent comparison with EKF and FKF Monte Carlo \acp{RMSE}. The optimizer minimizes $\mathrm{Tr}(\mathrm{PCRLB})$, but all values, figures, and tables present $\sqrt{\mathrm{Tr}(\mathrm{PCRLB})}$. Key simulation constants and optimization parameters used in this study are summarized in the Appendix. 

\subsection*{Case Study 1: Racetrack vs Optimal FFOV UAV Trajectory After Retasking}
\begin{figure}[!htb]
    \centering
    \includegraphics[width=0.5\textwidth]{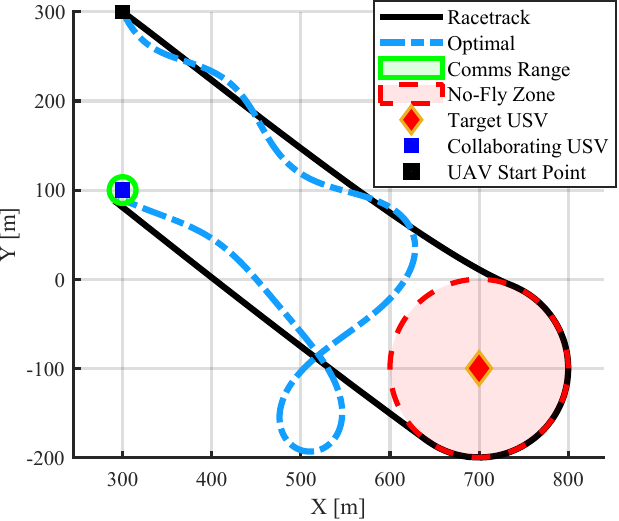}
    \caption{Heuristic UAV racetrack trajectory vs optimized Bernstein trajectory.}
    \label{fig:Racetrack_Trajectory}
\end{figure}

This case study compares two different localization strategies for a single \ac{FFOV} UAV following the unexpected discovery of a target during a broader maritime search mission as depicted in Fig. \ref{fig:Racetrack_Trajectory}. The scenario models a common operational context in which \acp{UAV} conduct generic surveillance until cued by external assets, such as manned platforms or distributed sensors, to localize an identified target. Following localization, the UAV must enter within a designated communication range to report the target's location relative to the collaborating USV.

At the moment of re-tasking, the UAV’s position is randomized within a bounded mission area to simulate the uncertainty inherent in real-world deployment. Localizing both USVs and reducing mission time are equally important. From this initial state, two localization strategies are considered.

The first approach utilizes a heuristic trajectory, commonly referred to as a “racetrack” pattern. This trajectory consists of an ellipsoid-like flight path designed to remain outside of the designated \ac{NFZ} while attempting to maintain continuous \ac{LOS} to the target. Upon re-tasking, the \ac{UAV} immediately enters the racetrack pattern to collect bearing-only measurements. While operationally straightforward, the racetrack approach is not designed to optimally localize a target while being time-efficient.

In contrast, the second approach employs Bernstein polynomial parameterization and numerical optimization to generate a feasible trajectory which minimizes the $\text{Tr(PCRLB)}$ for this mission. The trajectory is computed offline and explicitly incorporates vehicle kinematics, sensor \ac{FOV} constraints, and operational limits such as \ac{NFZ} boundary and the communication constraint. The resulting path adjusts the UAV's geometry to maximize bearing angle rate and maintain information-rich relative positions. While non-intuitive, this trajectory strategically exploits the UAV’s maneuverability to maximize information gain within the allowed sensing and operational constraints.

Each trajectory is evaluated in terms of the $\text{Tr(PCRLB)}$, which serves as the design metric. The \ac{EKF} is then separately employed in Monte Carlo trials to assess whether practical estimation performance approaches this theoretical lower bound. It is important to emphasize that the $\text{Tr(PCRLB)}$ is used as the design and evaluation metric, representing the best achievable performance for an estimator given the sensing geometry and measurement model. In contrast, the \ac{EKF} simulations reflect practical estimator performance, accounting for linearization errors and nonlinear dynamics. 

Table~\ref{tab:case_study1} summarizes the comparative results. The optimized trajectory achieves a lower $\sqrt{\mathrm{Tr}(\mathrm{PCRLB})}$ for the target and a comparable $\sqrt{\mathrm{Tr}(\mathrm{PCRLB})}$ for the collaborating USV, indicating improved information geometry for the target without a significant reduction in the measurements to the supporting platform. The EKF-based Monte Carlo validation further confirms these trends by showing an overall reduction in the standard deviation (SD), where the optimized trajectory achieves a 65.0\% reduction in target SD and a 12.8\% improvement in USV SD. Additionally, a 10.3\% reduction in total mission time relative to the racetrack baseline is achieved.

\begin{table}[!bp]
\centering
\caption{Comparison of Optimized and Heuristic FFOV UAV Performance}
\label{tab:case_study1}
\renewcommand{\arraystretch}{1.2}
\begin{tabular}{lcc}

\hline
{Metric} & {Optimized FFOV UAV} & {Heuristic FFOV UAV} \\
\hline
\multicolumn{3}{c}{{Target Data}} \\
\hline
$\sqrt{\mathrm{Tr}(\mathrm{PCRLB})}$, m & 0.93 & 2.08 \\
Standard deviation, m & 1.32 & 3.77 \\
\hline
\multicolumn{3}{c}{{Collaborating USV Data}} \\
\hline
$\sqrt{\mathrm{Tr}(\mathrm{PCRLB})}$, m & 0.70 & 0.64 \\
Standard deviation, m & 1.64 & 1.88 \\
\hline
\multicolumn{3}{c}{{Overall Simulation Data}} \\
\hline
Flight time, s & 47.3 & 52.7 \\
Total cost & 0.68 & 2.36 \\
\hline
\end{tabular}
\end{table}

\begin{figure}[!htb]
    \centering
    \includegraphics[width=.5\textwidth]{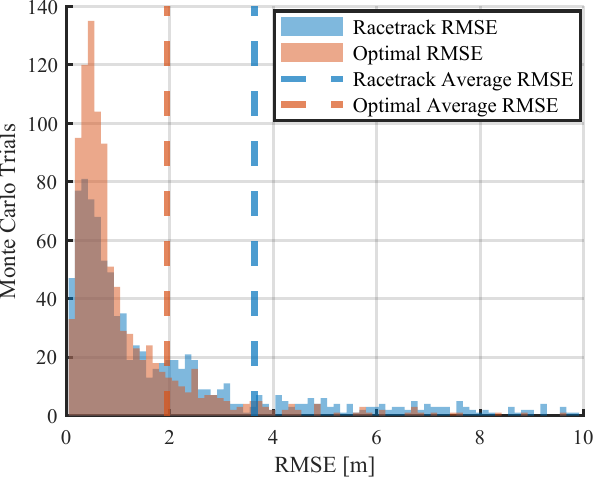}
    \caption{Histogram of combined target and collaborating-USV RMSE. The optimized trajectory reduces RMSE and outliers, whereas the racetrack trajectory exhibits higher variance and larger errors.}
    \label{fig:Case_Study_1_Histogram}
\end{figure}

The histogram in Fig.~\ref{fig:Case_Study_1_Histogram} illustrates the comparative distribution of combined RMSE values obtained using the heuristic and optimized trajectories. The vertical dashed lines denote the mean RMSE across all Monte Carlo runs. The heuristic trajectory results in a highly skewed RMSE distribution with a long tail, indicating occasional severe localization failures, whereas the optimized trajectory yields better performance based on the lower mean value compared to the heuristic trajectory.  Additionally, the optimized trajectory yields a significantly tighter RMSE distribution, implying better overall performance even in the worst-case scenario.

The improved $\sqrt{\mathrm{Tr}(\mathrm{PCRLB})}$ and SD performance of the optimized trajectory arises from how it distributes sensing opportunities over time. The racetrack trajectory focuses on the target USV during the first portion of the flight and transitions to localizing the collaborating USV in the second half. This results in highly informative and repeated observations of the target early on. However, since the estimator receives no new measurements of the target later in the mission, the final SD reflects accumulated process uncertainty. Conversely, the optimized trajectory continuously alternates between observing the target and the USV, preserving balanced observability and maintaining information flow until mission termination.

These results underscore the distinction between theoretical information metrics and realized estimator behavior. The $\mathrm{Tr}(\mathrm{PCRLB})$ provides a necessary but not sufficient indicator of achievable performance. Estimator accuracy ultimately depends on the temporal distribution and quality of measurements. The optimized trajectory demonstrates how an information-driven design can convert theoretical observability into consistent practical performance while reducing total mission time.

The single \ac{FFOV} UAV case study highlights both the benefits of trajectory optimization and the inherent limitations of \ac{FFOV} platforms. Without independent sensor steering, a single UAV cannot maintain optimal observation geometry over time, making estimation performance highly sensitive to vehicle dynamics and sensing alignment. To address these limitations, the subsequent case study extends the estimation framework to compare the target localization performance achieved by a team of coordinated \ac{FFOV} \acp{UAV} with that achieved by a single UAV equipped with a gimballed camera. Each FFOV UAV follows an optimized, complementary trajectory to maximize information gain. This comparison explores whether distributed coordination can compensate for individual sensor limitations and deliver comparable or superior localization performance. Additionally, the study examines broader system-level trade-offs, including resilience, scalability, and cost, which are factors critical to future operations in contested or resource-constrained environments.

\subsection*{Case Study 2: UAV equipped with a gimballed camera versus Multiple Fixed Field-of-View UAVs}
This case study evaluates the relative performance and system tradeoffs between a single UAV equipped with a gimballed camera and coordinated teams of \ac{FFOV} \acp{UAV} for bearing-only target localization. %
One or more \acp{UAV} are deployed from a collaborating \ac{USV} to localize a stationary adversarial surface target, with each \ac{UAV} operating at constant airspeed \(V_A\) and fixed altitude.

\begin{figure}[!hbtp]
    \centering
    \begin{subfigure}[t]{0.41\textwidth}
        \centering
        \includegraphics[width=\textwidth]{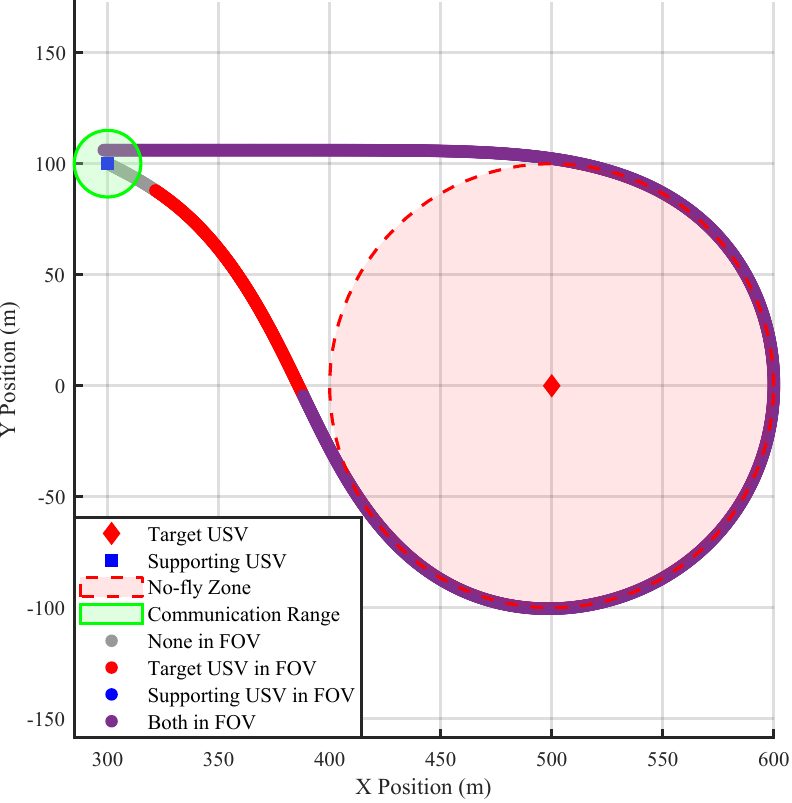}
        \caption{Optimized gimballed-UAV trajectory.}
        \label{fig:Optimal_Gimbal_FOV}
    \end{subfigure}
    \hfill
    \begin{subfigure}[t]{0.48\textwidth}
        \centering
        \includegraphics[width=\textwidth]{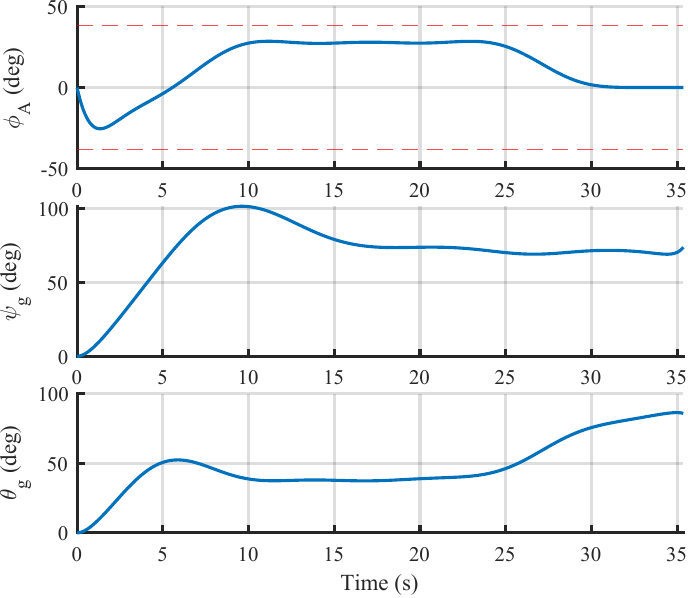}
        \caption{Gimbal control histories.}
        \label{fig:Gimbal_Controls}
    \end{subfigure}
    \caption{Optimized gimballed-UAV trajectory and corresponding control histories.}
    \label{fig:Gimbal_Combined}
\end{figure}

In the gimballed configuration, the \ac{UAV}'s sensor \ac{LOS} is decoupled  from the body-frame kinematics, enabling independent gimbal steering and preserving visibility regardless of flight direction. As shown in Fig.~\ref{fig:Optimal_Gimbal_FOV}, the trajectory is generated using a Bernstein polynomial-based optimization method that maximizes localization performance subject to mission constraints and the weighted time penalty in the cost function. %

The \ac{FOV} segmentation along the trajectory is color-coded to indicate intervals when the target vessel alone is visible (red), when the collaborating \ac{USV} alone is visible (blue), when both vessels are simultaneously visible (purple), and when neither vessel is visible (gray). For this optimized gimballed case, the trajectory maintains visibility of at least one vessel throughout the engagement, and portions of the path permit simultaneous observation of both vessels. This highlights the principal advantage of a gimballed architecture where sensor steering preserves measurement availability even when the airframe heading is constrained by vehicle dynamics. The corresponding bank-angle and gimbal-angle histories are shown in Fig.~\ref{fig:Gimbal_Controls}, where the red dashed lines denote the allowable bank-angle limits.

\begin{figure}[!thbp]
\centering
\includegraphics[width=0.48\textwidth]{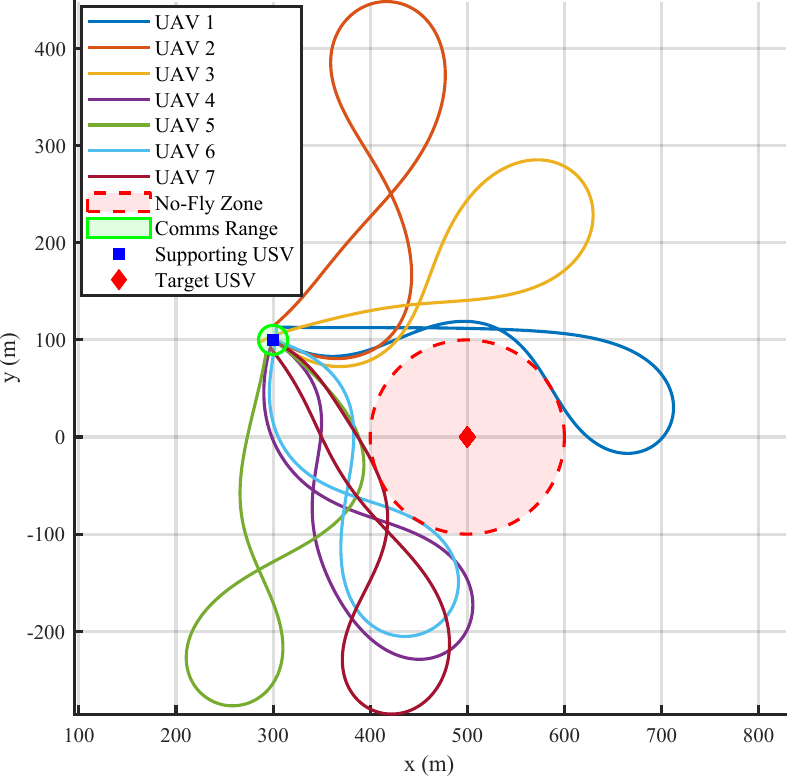}
\caption{Optimized trajectories for multiple FFOV \acp{UAV}. The \acp{UAV} localize a stationary target and a collaborating \ac{USV} in minimum time while adhering to mission flight constraints.}
\label{fig:FFOV_UAV_Traj}
\end{figure}

In contrast, \ac{FFOV} \acp{UAV} rely solely on airframe heading to align their sensors, requiring coordinated group behavior to maintain coverage. Fig.~\ref{fig:FFOV_UAV_Traj} shows trajectories for seven FFOV \acp{UAV} optimized collectively.

Table~\ref{tab:pcrlb_comparison} summarizes the optimization cost and corresponding lower bound on estimation accuracy for both the target and the collaborating \ac{USV}, and the final SD achieved in simulation. Since the optimization cost in Eq.~\ref{eq:cost_function} is a weighted sum of the target and collaborating \ac{USV} terms plus the time penalty, it represents a combined measure of estimation performance and mission duration.

A single \ac{FFOV} \ac{UAV} performs substantially worse than the gimballed configuration, with larger $\sqrt{\mathrm{Tr}(\mathrm{PCRLB})}$ values for both the target and collaborating \ac{USV}, longer mission time, higher optimization cost, and larger final target SD.  This gap underscores the sensing advantage provided by independent sensor articulation.  As the number of \ac{FFOV} \acp{UAV} increases, both estimation accuracy and overall cost improve. In general, the $\sqrt{\mathrm{Tr}(\mathrm{PCRLB})}$ values for the target and collaborating \ac{USV} decrease with cost, while the final target localization error follows the same overall trend. These improvements arise from increased geometric diversity and measurement redundancy as each additional \ac{UAV} contributes new information.

Although the FFOV solutions still require mission durations near the upper bound, the optimizer accepts this time expenditure because the resulting sensing geometry lowers localization error. By four to six FFOV \acp{UAV}, the team approaches the gimballed benchmark, and by seven \acp{UAV} it slightly outperforms the gimballed case in both $\sqrt{\mathrm{Tr}(\mathrm{PCRLB})}$ values, with a comparable optimization cost. The results also suggest diminishing returns, indicating that once sufficient angular diversity and measurement coverage are achieved, additional platforms provide progressively smaller gains in localization accuracy.

While the UAV equipped with a gimballed camera demonstrates strong localization performance in isolation, its advantages must be weighed against its higher \ac{SWaP-C} burden when assessing for scalability~\cite{Korobov_Shipitko_Konovalenko_Grigoryev_Chukalina_2020}. Gimballed systems require more power for stabilization, introduce mechanical complexity, and constitute a single point of failure. These characteristics make them less attractive for large-scale or resource-constrained operations where robustness, simplicity, and cost-effectiveness are highly valued~\cite{Petritoli_Leccese_Ciani_2018}.

\begin{table}[tb!]
\caption{\label{tab:pcrlb_comparison} $\mathbf{\sqrt{Tr({PCRLB})}}$, Optimization Cost, and Final Localization Error for the Target}
\centering
\small
\begin{tabular}{lccccccc}
\hline
 & & \multicolumn{2}{c}{$\sqrt{\mathrm{Tr}(\mathrm{PCRLB})}$, m} & Time to  & Optimization& Final \\
\cline{3-4}
UAV Type & Number  & Target & Collaborating USV &Complete, s&  Cost &Target SD, m \\
\hline
Gimballed & 1 & 1.580 & 1.708 & 35.96 & 2.0967 & 2.800 \\
\hline
FFOV & 1 & 3.833 & 1.775 & 44.23 & 13.1118 & 5.340 \\
     & 2 & 2.331 & 1.011 & 43.90 & 4.9939 & 2.694 \\
     & 3 & 2.272 & 0.887 & 45.00 & 4.7229 & 2.661 \\
     & 4 & 1.701 & 0.860 & 45.00 & 2.8217 & 2.023 \\
     & 5 & 1.612 & 0.823 & 45.00 & 2.5792 & 1.975 \\
     & 6 & 1.585 & 0.815 & 44.65 & 2.4943 & 1.651 \\
     & 7 & 1.430 & 0.789 & 44.73 & 2.1141 & 1.568 \\
\hline
\end{tabular}
\end{table}

Table~\ref{tab:uav_swa} provides an integrated comparison of performance and \ac{SWaP-C} characteristics. Notably, FFOV teams require more than a few platforms to approach the %
estimation performance of a UAV equipped with a gimballed camera, but larger coordinated teams can match or exceed that benchmark while offering advantages in scalability and resilience.

The data show that while a gimballed UAV achieves superior single-platform estimation performance, coordinated \ac{FFOV}  teams progressively recover that performance as additional \acp{UAV} are introduced. The single-\ac{FFOV}  case performs substantially worse than the gimballed benchmark in both $\sqrt{\mathrm{Tr}(\mathrm{PCRLB})}$ metrics, mission time, optimization cost, and final target SD, confirming the limitations imposed by rigid sensor mounting. However, as the number of \ac{FFOV}  \acp{UAV} increases, the estimation metrics improve steadily due to the geometric diversity and measurement redundancy created by coordinated sensing.

The larger \ac{FFOV} teams approach parity with the gimballed case. In particular, configurations with four to six \ac{FFOV} \acp{UAV} produce near-gimballed target performance, while the seven-\ac{FFOV}  configuration slightly surpasses the gimballed benchmark in both target and collaborating-USV $\sqrt{\mathrm{Tr}(\mathrm{PCRLB})}$ and yields a comparable optimization cost. Although the \ac{FFOV}  teams still require longer mission durations, these results show that coordinated multi-UAV systems can compensate for individual sensor limitations through sensing diversity alone.
\begin{table}[!tb]
\centering
\caption{\label{tab:uav_swa} Integrated Comparison of Estimation Performance and SWaP-C Metrics}
\small
\begin{tabular}{lcccccc}
\hline
 & & \multicolumn{2}{c}{$\sqrt{\mathrm{Tr}(\mathrm{PCRLB})}$} & Time to  & &  \\
\cline{3-4}
UAV Type & Number  & Target & Collaborating USV &Completion& SWaP-C & Resilience \\\hline

\hline
Gimballed & 1 & Baseline & Baseline & Baseline & High & None \\
\hline
FFOV & 1 & Poor & Poor & Poor & Low & None \\
     & 2 & Poor & Better & Poor & Low & Low \\
     & 3 & Near & Better & Poor & Low & Moderate \\
     & 4 & Near & Better & Poor & Low & High \\
     & 5 & Near & Better & Poor & Low & High \\
     & 6 & Near & Better & Poor & Low & High \\
     & 7 & Better & Better & Poor & Low & High \\
\hline
\end{tabular}
\end{table}
Operationally, this highlights the scalability and robustness of fixed-sensor swarms, where performance can be increased incrementally by adding additional low-cost \acp{UAV} rather than relying on a single high-capability platform. The FFOV architecture therefore offers superior resilience and scalability advantages. Its reduced \ac{SWaP-C} burden enables deployment in larger numbers, making it a compelling alternative for cost-sensitive or contested environments where attrition, endurance, or resource constraints are operational concerns.

\section{Conclusion}\label{Sec:Conclusion}
This paper investigated how trajectory optimization and sensor architecture affect bearing-only target localization using UAVs operating in GPS-denied environments. An information-theoretic optimization framework was developed to generate dynamically feasible trajectories using Bernstein polynomial parameterization, while an \ac{FKF} was used to fuse measurements and assess estimator performance.

The results showed that estimation-aware trajectory optimization substantially improves localization accuracy relative to heuristic flight paths, and that estimator performance approaches the theoretical \ac{PCRLB} limit. The results also showed that although a single gimballed UAV provides strong localization performance, coordinated teams of \ac{FFOV} UAVs can progressively recover and ultimately exceed that performance as additional platforms are introduced. This demonstrates that coordinated FFOV architectures can offset the sensing limitations of rigid body-mounted cameras through geometric diversity and measurement redundancy, while preserving advantages in \ac{SWaP-C}, resilience, and scalability.

The Bernstein polynomial parameterization played a critical role in enabling smooth, dynamically feasible trajectories that satisfied constraints such as NFZ avoidance, communication bounds, and platform limitations, while accounting for out-of-frame events. The study showed that trajectory optimization is not merely a control problem but a systems-level design tool that links estimation performance with practical mission considerations. These insights have direct implications for both tactical and commercial UAV operations. For military operations, FFOV UAV teams can enable targeting and over-the-horizon surveillance under SWaP-C and stealth constraints. In commercial applications, multi-agent FFOV systems may offer scalable, cost-effective solutions for inspection, mapping, and search-and-rescue missions. Overall, this work demonstrates that estimation-aware trajectory design enables scalable, resilient, and cost-effective UAV system designs for use in GPS-denied and contested environments.
 
\clearpage \FloatBarrier
\section*{Appendix: Simulation and Optimization Parameters}\label{Sec: Appendix}
\setcounter{table}{0}
\renewcommand{\thetable}{A\arabic{table}}

\begin{table}[htbp!]
\centering
\caption{Optimization weighting parameters for Case Studies 1 and 2}
\label{tab:opt_params_combined}
\renewcommand{\arraystretch}{1.1}
\footnotesize
\begin{tabular}{lcccc}
\hline
\multirow{2}{*}{Parameter} & \multirow{2}{*}{Symbol} &
\multicolumn{2}{c}{Value} \\
\cline{3-4}
 &  & Case Study 1 & Case Study 2 \\
\hline
Target weighting factor & $w_1$ & 0.5 & 0.9 \\
Collaborating USV weighting factor & $w_2$ & 0.5 & 0.1 \\
Time penalty weighting factor & $w_3$ & 0.5 & 0.9 \\
\hline
\end{tabular}
\end{table}

\begin{table}[htbp!]
\centering
\caption{Common simulation parameters for all UAV configurations}
\label{tab:sim_params_common}
\renewcommand{\arraystretch}{1.1}
\footnotesize
\begin{tabular}{lcc}
\hline
{Parameter} & {Symbol} & {Value} \\
\hline
No-fly zone radius & $r_{\text{NFZ}}$ & 100 m \\
Communication range & $r_{\text{com}}$ & 15 m \\
Initial time estimate & $t_{f}$ & 30 s \\
UAV speed & $V_{A}$ & 25 m/s \\
Average UAV altitude & z & 300 m\\
Maximum roll angle & $\phi^{\max}$ & $\pi/4$ rad \\
Camera sample rate & $f$ & 1 Hz \\
Measurement noise variance & $\sigma^2$ & $7.5\times10^{-6}$ rad$^2$ \\
Process noise variance & $\sigma_w^2$ & 0.7 \\
FFOV camera boresight angle & -- & $7\pi/18$ rad \\
Camera resolution (pixels) & -- & $2592 \times 1944$ \\
Camera focal length (mm) & -- & 2.0\\
\hline
\end{tabular}
\end{table}

\begin{table}[htbp!]
\centering
\footnotesize
\caption{Simulation parameters for the gimballed UAV configuration}
\label{tab:gimbal_sim_params}
\renewcommand{\arraystretch}{1.1}
\small
\begin{tabular}{lcc}
\hline
{Parameter} & {Symbol} & {Value} \\
\hline
Maximum azimuth angle & $\psi_{g}^{\max}$ & $\pi/2$ rad \\
Minimum azimuth angle & $\psi_{g}^{\min}$ & $-\pi/2$ rad \\
Maximum elevation angle & $\theta_{g}^{\max}$ & $\pi/2$ rad \\
Minimum elevation angle & $\theta_{g}^{\min}$ & $0$ rad \\
Maximum azimuth slew rate & $u_{\psi_{g}}^{\max}$ & 5.24 rad/s \\
Minimum azimuth slew rate & $u_{\psi_{g}}^{\min}$ & $-5.24$ rad/s \\
Maximum elevation slew rate & $u_{\theta_{g}}^{\max}$ & 5.24 rad/s \\
Minimum elevation slew rate & $u_{\theta_{g}}^{\min}$ & $-5.24$ rad/s \\
\hline
\end{tabular}
\end{table}

\FloatBarrier
\section*{Acknowledgments}
This work was funded in part by the Office of Naval Research under grant N0001425GI01578 and by the Consortium for Robotics and Unmanned Systems Education and Research (CRUSER) at the Naval Postgraduate School, Monterey, CA.

\bibliography{bibliography}

\end{document}